\documentclass[11pt,twoside]{article}
\usepackage{asp2004}
\usepackage{psfig}
\usepackage{epsf}
\usepackage{graphics}
\usepackage{lscape}
\usepackage{amssymb}
\def\edcomment#1{\iffalse\marginpar{\raggedright\sl#1\/}\else\relax\fi}
\reversemarginpar

\def\chandra{{\it CHANDRA}}

\def\ec{$\eta$ Carinae}
\def\rxte{{\it RXTE}}

\def\xmm{{\it XMM-Newton}}

\def\sax{{\it BeppoSAX}}

\def\mdunit{\mbox{$M_{\odot}$ yr$^{-1}$}}
\def\ks{~km s$^{-1}$}
\def\apj{ApJ}

\def\apjs{ApJS}
\def\apjl{ApJL}
\def\mnras{MNRAS}
\def\aap{A\&A}

\begin{document}
\title{
\begin{center}
X-ray Gyrations of \ec, \\
\textit{or} \\
 Is Presence of Evidence Evidence of Presence?
\end{center}
 }
\author{M.~F. Corcoran$^{1,2}$, K. Hamaguchi$^{1,3}$, T. Gull$^{4}$, A. Damineli$^{6}$, K. Davidson$^{6}$}
\affil{
1: NASA/GSFC/LHEA, Greenbelt, MD 20771; 
2: Universities Space Research Association, 7501 Forbes 
Blvd, Ste 206, Seabrook, MD 20706;
3: National Research Council, 500 Fifth Street, NW, Washington, D.C. 20001;
4: NASA/GSFC/LASP, Greenbelt, MD 20771;  
5: Instituto Astron{\^ o}mico e Geof{\' i}sico da USP, R. do Matao 1226,
05508-900 S\~ao Paulo
Brazil; 
6: Astronomy Department, University of Minnesota, 116 Church Street SE, Minneapolis,
MN 55455
}

\begin{abstract}
We review the properties of the variable X-ray emission from the extremely massive star \ec\ concentrating on the last X-ray minimum, and briefly consider the possible role of a binary companion star on the observational properties of the system.
\end{abstract}
\thispagestyle{plain}

\section{Variable X-rays from \ec\  -- Signs of a Companion Star?}

The enigmatic star \ec\ is extremely luminous and believed to be very massive ($\sim 100 M_{\odot}$) and to lie very near the Eddington Limit.  It
serves as an example of a possible hypernova/Gamma-ray burst precursor \citep{heger03} and  as a crude example of the supermassive stellar objects thought to form first in the Universe \citep{abel98}.

Recent evidence  \citep{pw94, dam96, dunc95,corc95} suggests strongly that \ec\ is  a binary system with a 5.54-year period. Continued monitoring at optical, radio and X-ray wavelengths has shown that the emission in these wavebands is strongly correlated.  Every 5.5 years the radio, IR, optical and X-ray emission all  experience a brief minimum in intensity.

\begin{figure}
  \begin{center}
    \plotone{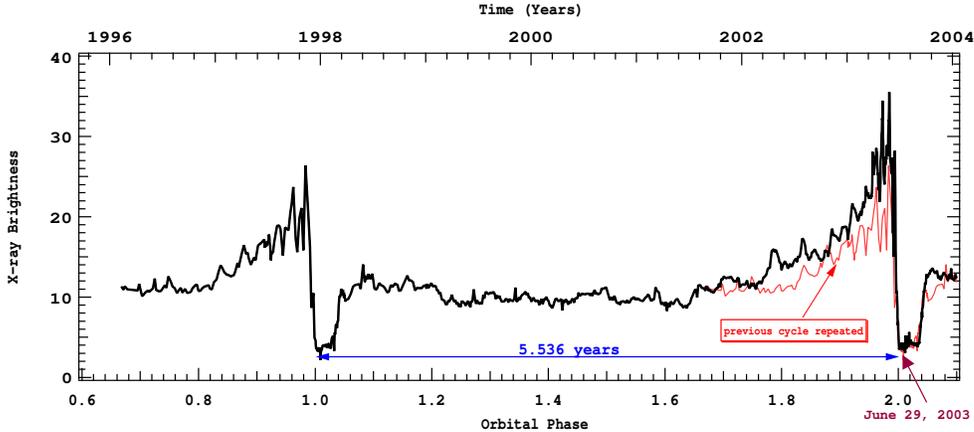}
        \caption{\small $2-10$ keV X-ray brightness of \ec\ as measured by the \rxte\ satellite from 1996-2003.  The X-ray ``lightcurve'' shows a gradual increase in X-ray brightness prior to the decline to a minimum which lasts 3 months.}
    \label{fig:lc_orbit}
  \end{center}
\end{figure}

Figure \ref{fig:lc_orbit} shows the X-ray brightness in the $2-10$ keV band since 1996 as observed by the Rossi X-ray Timing Explorer (\rxte).  The start of X-ray minima is given by 
\begin{equation}
1997.955 + 5.536E,
\end{equation}
where $E$ is the cycle count and the epoch is derived from daily monitoring observations with \rxte\ prior to the 1998 X-ray minimum.  The period, $P=5.536$ years, is the interval between the start of the consecutive minima.  Given the \rxte\ sampling near the minima, the nominal uncertainty on the period is less than one day.  

%Figure \ref{fig:nh} shows the column densities, $N_{H}$, derived by fitting spatially-resolved X-ray spectra obtained by  \asca, \sax, \chandra, \& \xmm\ around the time of the X-ray minima in 1998 \& 2003.  The column densities are clearly larger during the minimum than prior to the minimum.  In addition the interval of enhanced absorption extends beyond the ``recovery'' of the X-ray flux. 

%\begin{figure}[!ht]
%  \begin{center}
%    \plotone
%% Other \includegraphics options include: height, width, angle, scale
%    \caption{\small The radio lightcurve of \ec\ from Duncan \& White from 1992--2004 (see http://www.astro.umd.edu/$\sim$white/images/eta\_time\_full.html). }
%    \label{fig:ecradio}
%  \end{center}
%\end{figure}

The X-ray emission is believed to arise from a wind-wind collision.  The periodic behavior of the X-ray emission is thought to be a consequence of orbital eccentricity coupled with variation in the amount of absorbing material in front of the colliding wind shock. The maximum X-ray temperature is about 50 million K, suggesting that the companion's wind velocity is $\sim$ 3000\ks\ 
\citep[much higher than the measured wind speed of \ec\ itself, $\approx 500$\ks,][]{hill01}. 
The mass loss rates from the X-ray spectra are $\dot{M}_{\eta} \approx 10^{-4}$\mdunit\ for \ec\ (smaller than the mass loss rates derived from radio and millimeter observations) and $\dot{M}_{c} \approx 10^{-5}$\mdunit\ for the companion star \citep{jmp02b}.

%\begin{figure}[htbp] %  figure placement: here, top, bottom, or page
%   \begin{center}
%   \plotone{orbit.eps} 
%   \caption{\small Orbit of the companion star around \ec\ based on the orbital elements from \citet{corc01}.}
%   \label{fig:orbit}
%   \end{center}
%\end{figure}

%\begin{figure}[htbp] %  figure placement: here, top, bottom, or page
%   \centering
%   \includegraphics[width=3in]{orbit.eps} 
%   \caption{\small Orbit of the companion star around \ec\ based on the orbital elements from \citet{corc01}.}
%   \label{fig:orbit}
%\end{figure}

There are discrepancies between the observed X-ray emission and the colliding wind models:
1) the X-ray flux is expected to be strictly periodic, yet the observed emission shows significant cycle-to-cycle variations (though these discrepancies have diminished in the months following the X-ray minimum\footnote{
see 
http://lheawww.gsfc.nasa.gov/users/corcoran/eta\_car/etacar\_rxte\_lightcurve/ 
}); 2) the X-ray
emission is expected to be more symmetric  around periastron \citep{jmp98}, yet the X-ray brightness prior to eclipse ingress is about a factor of  3 higher than the brightness after recovery from the eclipse (which, however, is similar to the asymmetry around the X-ray minimum in the colliding wind binary WR 140); 3) unanticipated  variations or ``spikes'' occur on a timescale of $\sim 80-100$ days \citep{bish97, corc97, kd98}.   

\section{Spectral and Spatial X-ray Variations}

An observing campaign with the \chandra\ X-ray Observatory obtained 5 observations with the high-energy X-ray transmission gratings prior to, during and after the 2003 X-ray minimum.  During the minimum, residual emission at zeroth order was clearly detected from \ec, along with spatially-resolved hard X-ray emission associated with the Homunculus Nebula around \ec.  This hard emission is believed to be the X-ray reflection of the time-delayed X-ray flux from the colliding-wind source produced by electron-scattering from material in the Homunculus \citep{corc04}.  The dispersed spectra of the colliding wind source shows dramatic changes which, as far as we are aware, have never been reported for any other non-degenerate source.  We see variations in the relative strengths of the forbidden to intercombination ($f/i$) line ratios in He-like Si and S.  Additionally, the Si and S line centroids show increasingly  negative radial velocities near the X-ray minimum.  The variation of the Si and S line centroids is  similar in some respects to the radial velocity shift of the He II 4686 line reported by \citet{sd04}.  Interestingly, the Fe XXV blend is redshifted, and grows more redshifted prior to the X-ray minimum, in contrast to the behavior of the Si and S lines.  In addition,  there is an apparent excess of emission, or perhaps a low-energy ``tail'', visible between the Fe XXV blend and the Fe fluorescent line in some of the spectra.  This ``tail'' is most apparent in the grating observation on June 16, 2003, i.e. the last grating observation before the X-ray minimum. This ``tail'' can also be seen in \xmm\ observations during the minimum (as discussed by Hamaguchi elsewhere in this volume).  The \chandra\ spectra and the \xmm\ spectra show clearly that the absorbing column to the colliding wind source increases during the minimum and this absorption enhancement continues even after the end of the X-ray minimum \citep[a result first seen in \sax\ spectra during the minimum of 1998,][] {viotti02}.

\section{The Role of a Companion on the Nuclear and Dynamical Evolution of \ec}

A key question is what role the companion plays in the evolution of \ec\ in particular, and what role binarity plays in the evolution of extremely massive stars in general. Formation of extremely massive stars ($\sim 100M_{\odot}$) via competitive accretion of lower-mass stars ($\sim10M_{\odot}$) in a dynamical collapse phase of a young cluster is one way to overcome radiative and angular momentum barriers \citep{bon04} and some simulations have shown this process to result in creation of an extremely massive star orbited by a lower mass companion in a long-period elliptical orbit \citep{bon02}.   
The evolution of binary systems can differ from that of single stars due to  exchanges of mass and/or angular momentum in the system. For example, transfer of orbital angular momentum to rotational angular momentum of the primary could presumably cause large instabilities by reducing the effective gravity of  the primary, driving it closer to the Eddington Limit.

\begin{figure}[!ht]
  \begin{center}
   \plotone{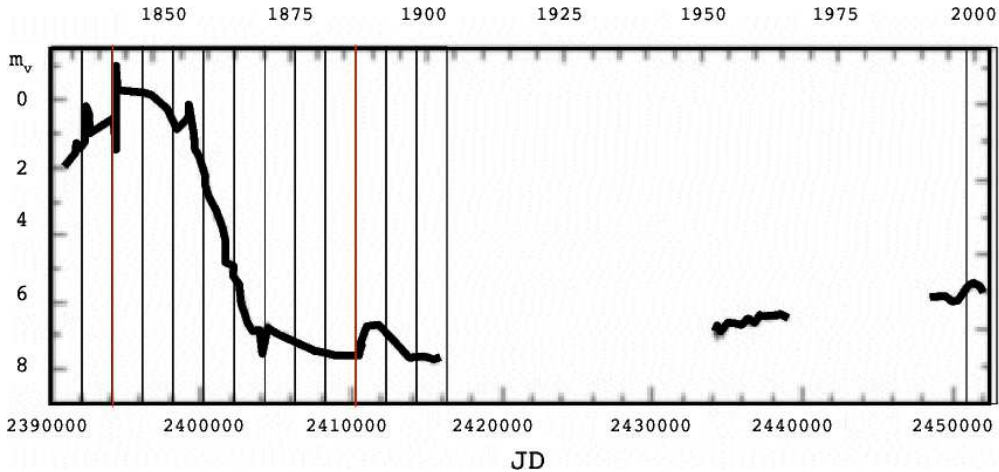}
\caption{\small Times of X-ray minima compared to the historical V-band lightcurve. The start of the rapid brightenings of 1838 and 1843 (the ``Great Eruption'') apparently were associated with X-ray minima.  In addition the 1890 ``mini-eruption'' was also apparently associated with an X-ray minimum, as were other times of rapid brightness variations. }
  \label{fig:xraytiming}
  \end{center}
\end{figure}

It is unclear what, if any, role the companion plays in the evolution of \ec.  
\ec\ has undergone at least two major eruptions since the 1830's and it's an interesting question how the timing of  these events relate to the X-ray minima. 
As noted in \citet{dam96} and by David Frew  there is some circumstantial evidence that the timings of ``spectroscopic events'' are associated with these outbursts.  Figure \ref{fig:xraytiming} shows the historical light curve of \ec\ \citep{cs00} with times of X-ray minima (based on the ephemeris given above) marked by vertical lines. X-ray minima are believed to be associated with periastron passages, so correlations between times of X-ray minima and large scale eruptions suggest a physical interaction between the two stars when the stars are close.  It's already been suggested that the close approach of the companion causes enhanced mass loss from \ec\ \citep{corc01} to almost completely obscure the X-ray emitting region.  The apparent increase in the X-ray absorbing column after the start of the X-ray minimum in 2003 provides some support to this assertion.  Perhaps the large eruptions of the nineteenth century were simply scaled-up versions of episodes which occur every periastron pass. It can be speculated that such periodic ejections of matter, if sufficiently optically thick, may mimic the characteristics of the ``shell events'' \citep{zanella84} which have been attributed to \ec. On the other hand, simple dynamical simulations suggest that the passage of the companion near the primary may distort the wind of the primary which could in principle produce an increase in density along the line of sight to the X-ray emitting colliding-wind shock, which would mimic an enhancement in the mass-loss rate from \ec\ \citep[see, for eg.,][]{dfg04}.  

The companion's radiation field  should also have a significant effect on \ec's mass loss, since from apastron to periastron the radiative flux from the companion as seen by \ec\ varies by about a factor $>200$.  The periastron passage of the companion around \ec\ might cause significant radiative heating of the photosphere or inner wind of \ec, which might significantly alter the amount of mass lost from the heated regions of the star.  Another effect which may play a role is ``sudden radiative braking'' \citep{kg97}  in which the radiative flux from \ec\ might slow the wind from the companion to moderate the shock when the stars are close.  However the analysis of the \xmm\ spectra by Hamaguchi (these proceedings) suggests little change in the shape of the high-temperature emission, which suggests little change in the velocity of the companion's wind during the X-ray minimum. 

The ejection of significant amounts of material from a binary should result in significant dynamical evolution due to the loss of angular momentum from the system.  If the correlation between the X-ray minima times and the times of the start of the large eruptions is physical, then this correlation argues that the period of the system has remained $\sim$constant for over 150 years.  This also suggests that the amount of matter and angular momentum lost was only a small fraction of the system total. If true this would severely constrain models of the mass and angular momentum in the Homunculus Nebula, and would appear to rule out models in which the material in the Homunculus originated from an eruption of the companion star \citep[for e.g.,][]{lamers}.

%\section*{Appendix~E: Journal Abbreviations}
%\begin{center}
%\begin{tabular}{ll}
%\verb"\aj" & Astronomical Journal (\aj )\\
%\verb"\araa" & Annual Review of Astronomy and Astrophysics(\araa )\\
%\verb"\apj" & Astrophysical Journal (\apj )\\
%\verb"\apjs" & \rule[.5ex]{2em}{.4pt}, Supplement Series (\apjs )\\
%\verb"\ao" & Applied Optics (\ao )\\
%\verb"\apss" & Astrophysics and Space Science(\apss )\\
%\verb"\aap" & Astronomy and Astrophysics (\aap )\\
%\verb"\aaps" & \rule[.5ex]{2em}{.4pt}, Supplement Series (\aaps )\\
%\verb"\azh" & Astronomicheskii Zhurnal (\azh )\\
%\verb"\baas" & Bulletin of the AAS (\baas )\\
%\verb"\jrasc" & Journal of the RAS of Canada (\jrasc )\\
%\verb"\memras" & Memoirs of the RAS (\memras )\\
%\verb"\mnras" & Monthly Notices of the RAS (\mnras )\\
%\verb"\nat" & Nature (\nat )\\
%\verb"\pra" & Physical Review A: General Physics (\pra )\\
%\verb"\prb" & Physical Review B: Solid State (\prb )\\
%\verb"\prc" & Physical Review C (\prc )\\
%\verb"\prd" & Physical Review D (\prd )\\
%\verb"\prl" & Physical Review Letters (\prl )\\
%\verb"\pasp" & Publications of the ASP (\pasp )\\
%\verb"\pasj" & Publications of the ASJ (\pasj )\\
%\verb"\qjras" & Quarterly Journal of the RAS (\qjras )\\
%\verb"\science" & Science (\science )\\
%\verb"\skytel" & Sky and Telescope (\skytel )\\
%\verb"\sovast" & Soviet Astronomy (\sovast )\\
%\verb"\ssr" & Space Science Reviews (\ssr )\\
%\verb"\zap" & Zeitschrift f\"ur Astrophysik (\zap )\\
%\end{tabular}
%\end{center}

\end{document}